\documentclass[twocolumn]{aastex62}

\graphicspath{{./}{figures/}}

\received{February 25, 2020}
\revised{March 18, 2020}
\accepted{March 18, 2020}

\submitjournal{ApJL}

\shortauthors{Bennet et al.}

\begin{document}

\title{The Satellite Luminosity Function of M101 into the Ultra-Faint Dwarf Galaxy Regime}

\correspondingauthor{Paul Bennet}
\email{paul.bennet@ttu.edu}

\author[0000-0001-8354-7279]{P. Bennet}
\affiliation{Physics \& Astronomy Department, Texas Tech University, Box 41051, Lubbock, TX 79409-1051, USA}
\author[0000-0003-4102-380X]{D. J. Sand}
\affiliation{Steward Observatory, University of Arizona, 933 North Cherry Avenue, Rm. N204, Tucson, AZ 85721-0065, USA}
\author[0000-0002-1763-4128]{D. Crnojevi\'c}
\affiliation{University of Tampa, 401 West Kennedy Boulevard, Tampa, FL 33606, USA}
\author[0000-0002-0956-7949]{K. Spekkens}
\affiliation{Department of Physics and Space Science, Royal Military College of Canada P.O. Box 17000, Station Forces Kingston, ON K7K 7B4, Canada}
\affiliation{Department of Physics, Engineering Physics and Astronomy, Queen’s University, Kingston, ON K7L 3N6, Canada}
\author[0000-0001-8855-3635]{A. Karunakaran}
\affiliation{Department of Physics, Engineering Physics and Astronomy, Queen’s University, Kingston, ON K7L 3N6, Canada}
\author[0000-0002-5177-727X]{D. Zaritsky}
\affiliation{Steward Observatory, University of Arizona, 933 North Cherry Avenue, Rm. N204, Tucson, AZ 85721-0065, USA}
\author[0000-0001-9649-4815]{B. Mutlu-Pakdil}
\affiliation{Steward Observatory, University of Arizona, 933 North Cherry Avenue, Rm. N204, Tucson, AZ 85721-0065, USA}

\begin{abstract}

We have obtained deep {\it Hubble Space Telescope (HST)} imaging of four faint and ultra-faint dwarf galaxy candidates in the vicinity of M101 -- Dw21, Dw22, Dw23 and Dw35, originally discovered by \citet{bennet17}. Previous distance estimates using the surface brightness fluctuation technique have suggested that these four dwarf candidates are the only remaining viable M101 satellites identified in ground based imaging out to the virial radius of M101 ($D$$\approx$250 kpc).  Advanced Camera for Surveys imaging of all four dwarf candidates shows no associated resolved stellar populations, indicating that they are thus background galaxies.  We confirm this by generating simulated {\it HST} color magnitude diagrams of similar brightness dwarfs at the distance of M101. Our targets would have displayed clear, resolved red giant branches with dozens of stars if they had been associated with M101.  With this information, we construct a satellite luminosity function for M101, which is 90\% complete to $M_V$=$-$7.7 mag and 50\% complete to $M_V$=$-$7.4 mag, that extends into the ultra-faint dwarf galaxy regime.  The M101 system is remarkably poor in satellites in comparison to the Milky Way and M31, with only eight satellites down to an absolute magnitude of $M_V$=$-$7.7 mag, compared to the 14 and 26 seen in the Milky Way and M31, respectively. Further observations of Milky Way analogs are needed to understand the halo-to-halo scatter in their faint satellite systems, and connect them with expectations from cosmological simulations.

\end{abstract}


\section{Introduction} \label{sec:intro}

The faint end of the satellite luminosity function is an important testing ground for the $\Lambda$ Cold Dark Matter ($\Lambda$CDM) model for structure formation \citep[e.g.][]{Planck18}, and for understanding how galaxies form in the smallest dark matter halos.  Despite many successes, challenges remain in reproducing the number, structure, luminosity and distribution of faint dwarf galaxy satellites around their larger hosts \citep[see][for a recent review]{Bullock17}. Most effort has been focused on reproducing the satellite systems of the Milky Way (MW) and M31 \citep[see e.g.][for recent results]{Drlica-Wagner19} into the `ultra-faint' dwarf galaxy regime ($M_V$$\gtrsim$$-$7.7, or L$\lesssim$10$^5$ L$_{\odot}$, using the definition of \citealt{Simon19}).

The faint satellite luminosity function of nearby galaxy systems adds context to Local Group studies, and illustrates how the satellite luminosity function changes with primary halo mass, environment and morphology.  For these reasons, several wide-field imaging and spectroscopic surveys of nearby galaxy systems have been initiated, across a range of central galaxy masses  \citep[e.g.][]{chiboucas13,Sand14,Sand15a,crnojevic14b,Muller15,crnojevic16,carlin16,toloba16,bennet17,Smercina17,Geha17,Smercina18,Crnojevic19, carlsten19_many, Muller19,Bennet19}.

One focus of faint satellite galaxy studies beyond the Local Group has been M101, which has a stellar mass similar to that of the MW ($\sim$5.3$\times$10$^{10}$ M$_{\odot}$; \citealt{vanDokkum14}), and is at a distance of  $D$=6.52$\pm$0.19 Mpc \citep[which we will use throughout this work;][]{Beaton19}  
amenable to efficient {\it Hubble Space Telescope (HST)} follow-up. This {\it HST} follow up is essential, as dwarf galaxy candidates at this distance can be identified by their diffuse stellar light from the ground, but to confirm their association with M101 requires resolving the dwarf's stars and measuring a tip of the red giant branch (TRGB) distance.   For M101, {\it HST} follow-up of many dwarf candidates has resulted in associations with the background galaxy group, NGC~5485 ($D$$\sim$27 Mpc; \citealt{Merritt16}), and its presence has complicated interpretations of candidates from ground-based data alone.  

Recent searches for M101 dwarfs started with the Dragonfly survey \citep{merritt14}, which ultimately uncovered three new M101 satellites with {\it HST}-derived TRGB distance measurements \citep[M101 DF1, M101 DF2, and M101 DF3;][]{Danieli17}.  Other teams identified further diffuse dwarf candidates from the ground \citep{karachentsev15,Java16,muller17}, while a comprehensive, semi-automated search using data from the Canada-France-Hawaii Telescope (CFHT) Legacy Survey identified 39 additional, new candidates \citep{bennet17}.  Taking this collection of M101 diffuse dwarf galaxy candidates, \citet{carlsten19} applied a new calibration of the surface brightness fluctuation (SBF) distance measuring technique \citep{carlsten19a}. Out of the 43 identified dwarf candidates found by other groups, Carlsten et al. identified 2 that were very likely to be associated with M101 (DwA and Dw9), with a further 12 whose distance uncertainties also made them possible candidates.  Follow-up {\it HST} imaging of 19 dwarf galaxy candidates confirmed that DwA and Dw9 are M101 group members, verified by their TRGB distance, and that the remainder of their sample are all background objects \citep{Bennet19}.  Using the collected M101 dataset, \citet{Bennet19} constructed a satellite luminosity function for M101 that is complete to $M_V$$\approx$$-$8, and showed that M101 has a very sparse satellite population in contrast to the MW and M31.  Further, \citet{Bennet19} speculated that this may be due to the relative isolation of M101, as a comparable system with few satellites, M94, has a similarly isolated environment \citep{Smercina18}.  The relative isolation of M101 is defined using its tidal index ($\Theta_{5}$) from \cite{Karachentsev13}. This quantity uses the magnitude of tidal force exerted on a galaxy by its five most influential neighbours as a proxy for environment. The tidal index is then normalized such that zero indicates an isolated galaxy. 
With tidal indices of 0.5 and -0.1 respectively we consider M101 and M94 to be relatively isolated compared to other Local Volume hosts such as M31 ($\Theta_{5}$$=$1.8) or M81 ($\Theta_{5}$$=$2.6) \citep{Karachentsev13}. 
Follow-up HI observations of a large sample of M101 dwarf candidates confirmed that several were associated with the background galaxy group NGC~5485 via velocity measurements, and that the faintest M101 satellites within the virial radius of M101 are quenched just as those in the Local Group \citep{K20}.

Here we present {\it Hubble Space Telescope} follow-up imaging of the four remaining viable M101 dwarf candidates which are not ruled out as satellites by SBF-derived distance limits -- Dw21, Dw22, Dw23, and Dw35 -- all of which were found by a semi-automated dwarf detection algorithm \citep{bennet17}, but were not previously observed in the {\it HST} study of \citet{Bennet19}.  These four satellites are very faint, and at the distance of M101 would correspond to absolute magnitudes between $M_V$=$-$7.4 and $M_V$=$-$8.1.  Thus, by confirming their identity as either M101 satellites or background objects, we can extend the luminosity function of M101 to $M_V$$\approx$$-$7.4 well into the ultra-faint dwarf galaxy regime. This is crucial to measure the dispersion of satellite properties as a function of mass and environment, and for continuing comparisons with the Local Group.

\section{HST Data and Photometry} \label{sec:data}

\floattable
\begin{deluxetable}{c|cc|cccc|cc}
\tablecaption{Unresolved Dwarf candidates \label{tab:obj_unres}}
\tablehead{
\colhead{Name} & \colhead{RA} & \colhead{Dec} & \colhead{V-band} & \colhead{V-band} & \colhead{F606W} & \colhead{F814W} & \colhead{Half light} & \colhead{Half light}\\
\colhead{} & \colhead{} & \colhead{} & \colhead{Magnitude} & \colhead{Magnitude} &\colhead{Magnitude} & \colhead{Magnitude} & \colhead{radius (CFHTLS)} & \colhead{radius (HST)}\\
\colhead{} & \colhead{} & \colhead{} & \colhead{(CFHTLS)} & \colhead{(HST)} & \colhead{(HST)} &\colhead{(HST)} & \colhead{(arcsec)} & \colhead{(arcsec)\tablenotemark{a}}}
\colnumbers
\startdata
Dw21 & 14:07:56.5 & +54:56:03 & 21.2$\pm$0.2 & 21.2$\pm$0.4 & 21.4$\pm$0.4 & 21.2$\pm$0.5 & 3.26$\pm$0.74 & 3.81$\pm$1.25\\
Dw22 & 14:03:03.3 & +54:47:12 & 21.0$\pm$0.1 & 20.8$\pm$0.3 & 21.0$\pm$0.3 & 20.5$\pm$0.3 & 3.41$\pm$0.31 & 2.89$\pm$0.60\\
Dw23 & 14:07:08.4 & +54:33:49 & 21.2$\pm$0.5 & 21.7$\pm$0.5 & 21.9$\pm$0.5 & 21.5$\pm$0.2 & 9.30$\pm$7.60 & 2.87$\pm$0.89\\
Dw35 & 14:05:36.2 & +54:49:02 & 21.8$\pm$0.2 & 21.6$\pm$0.3 & 21.8$\pm$0.3 & 21.3$\pm$0.3 & 2.62$\pm$0.58 & 2.14$\pm$0.34\\
\enddata
\tablenotetext{a}{Derived from F606W images}
\tablenotetext{}{ NOTE -- col(1): Candidate name. col(2) \& col(3): J2000 position of optical centroid. col(4): V-band magnitude, based on CFHTLS imaging \citep{bennet17}. col(5): V-band magnitude, based on the F606W {\it HST} imaging, converted via the relation from \cite{Sahu14}. col(6) \& col(7): F606W and F814W magnitude, based on {\it HST} imaging. col(8): The half-light radius of the candidates, based on CFHTLS imaging \citep{bennet17}. col(9): The half-light radius of the candidates, based on HST imaging. }
\end{deluxetable}

We obtained {\it HST} images (GO-15858; PI: Bennet) of four remaining M101 dwarf candidates from the \cite{bennet17} sample, identified as viable M101 satellites by \citet{carlsten19} based on their SBF-derived distances.  The data were obtained using the Wide Field Camera of the Advanced Camera for Surveys (ACS), with each dwarf placed on a single ACS chip.  Each target was observed for one orbit, split between the F606W and F814W filters, with an exposure time of $\sim$1000-1200 seconds per filter.

We perform PSF-fitting point source photometry on the ACS images as described in \citet{Bennet19}, which we briefly describe here.  We use the DOLPHOT v2.0 photometric package \citep{dolphin00}, and the suggested input parameters from the DOLPHOT User's Guide\footnote{\url{http://americano.dolphinsim.com/dolphot/dolphotACS.pdf}}.  
Standard photometric quality cuts are then applied using the following criteria: the derived photometric errors must be $\leq$0.3 mag in both bands, the sum of the crowding parameter in both bands is $\leq$1 and the squared sum of the sharpness parameter is $\leq$0.075. Detailed descriptions of these parameters can be found in \cite{dolphin00}. Further, extensive artificial star tests were performed to assess our photometric errors and completeness.  The 50\% completeness limits for F814W and F606W are $\sim$26.8 and $\sim$27.5 mag, respectively, across all HST images. 
The derived magnitudes for each point source were corrected for foreground MW extinction using the \citet{Schlafly11} calibration of the \citet{schlegel98} dust maps.

\section{The Nature of the Dwarf Candidates}

\begin{figure*}
 \begin{center}
 \includegraphics[width=16cm]{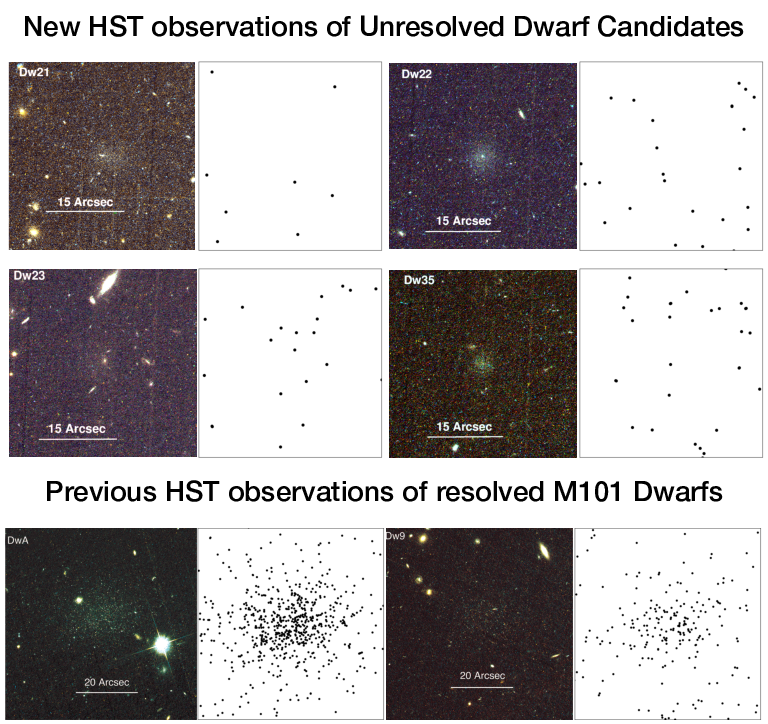}
 \caption{In the top set of four panels we show colorized {\it HST} cutouts of the four M101 dwarf galaxy candidates presented in the current work: Dw21, Dw22, Dw23 and Dw35.  Alongside each color image we show spatial plots of all point sources found by DOLPHOT after quality cuts.  There is no resolved stellar overdensity at the position of any of the newly targeted dwarf candidates.  Images are 0.6' $\times$ 0.6' for the new candidates; north is up and east is to the left.  For contrast, in the bottom set of panels we show colorized {\it HST} cutouts from two confirmed M101 satellites presented in \citet{Bennet19} -- DwA (M$_V$$=$$-$9.5) and Dw9 (M$_V$$=$$-$8.2); these images are 1.0' $\times$ 1.0' due to the larger size of these objects.  In this case, each dwarf shows a clear, associated point source overdensity, indicating that we are resolving it into stars. \label{fig:diffuse}}
 \end{center}
\end{figure*}

Inspection of the point-source photometry for our four M101 dwarf candidates reveals no associated resolved stellar overdensities, and only diffuse emission is observed, as we illustrate in Figure~\ref{fig:diffuse}. This is in contrast to known M101 satellites, which are resolved into stars with similar {\it HST} observations (see Figure \ref{fig:diffuse}, bottom panels).  This indicates that the individual stars that make up the TRGB are too faint to be detected in our {\it HST} imaging, and therefore that these dwarf candidates are in the background. We investigate the expected CMDs of our targets if they were actually associated with M101 below. 

Assuming a luminosity of $M_{I}^{TRGB}$$\approx$$-$4 mag for the TRGB \citep[see, for instance,][]{Gallart05,Radburn11}, and our measured 50\% completeness limit of $F814W$=26.8 mag, an undetected TRGB implies a distance modulus $\gtrsim$30.8 mag, corresponding to a distance $\gtrsim$14.5 Mpc -- well beyond the distance of M101. These dwarfs are potentially members of the NGC 5485 group as they project within that group's virial radius \citep{K20}, however we can not definitively state whether or not they are members of the NGC 5485 group from these observations.  

\begin{figure*}
 \begin{center}
 \includegraphics[width=6.8cm]{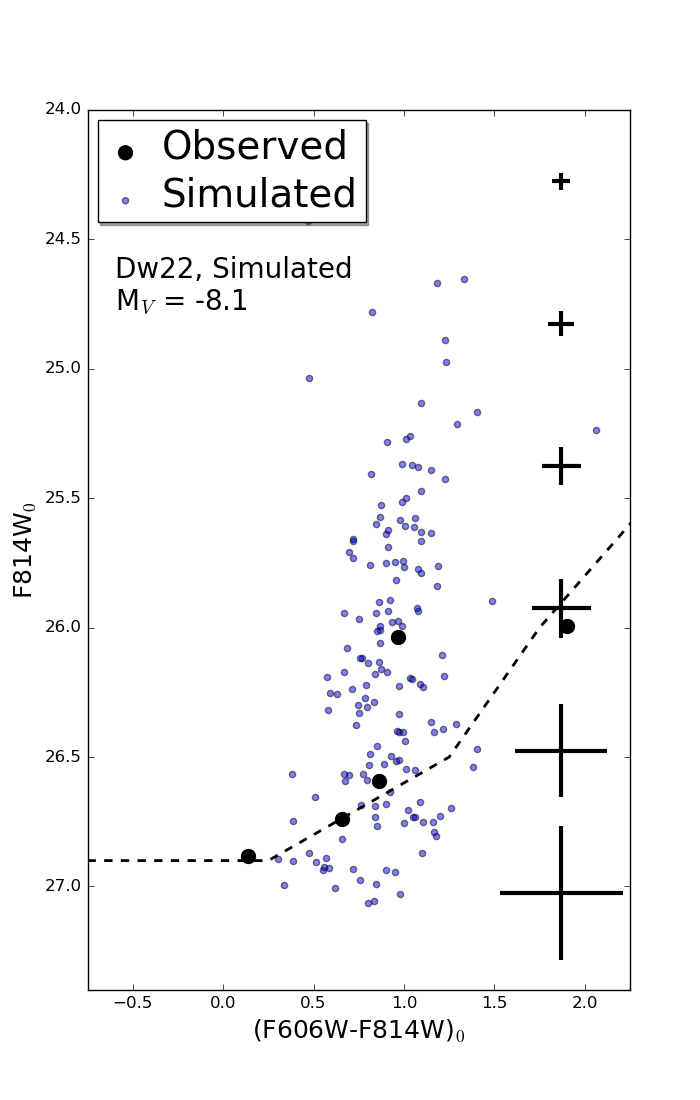}
 \includegraphics[width=6.8cm]{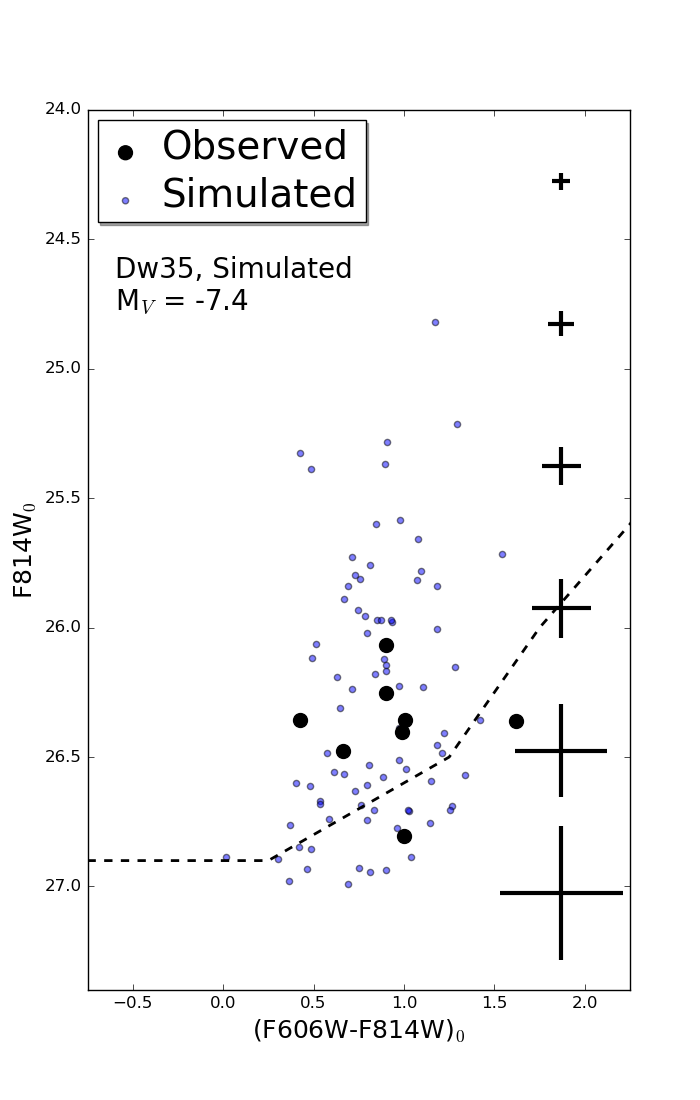}
 \caption{The measured CMDs for resolved sources (large black points) in two of our four M101 dwarf candidates, Dw22 and Dw35  
 -- these two objects were chosen to bound the brightness range of our sample. The lower dashed line indicates the photometric 50\% completeness limit; photometric uncertainties are shown along each CMD. 
 Very few point sources are found to be associated with each object.  Based on their ground-based brightness, we have simulated expected CMDs given our measured completeness and photometric uncertainties, and assuming each object is at the distance to M101 (small blue points).  Given the large differences between our measured CMDs and the expectations based on their ground-based brightness (which we have confirmed with our {\it HST}-based GALFIT measurements), we conclude that none of the four dwarf candidates are associated with M101, and are instead background objects. 
 \label{fig:CMD}}
 \end{center}
\end{figure*}

To further illustrate the distant nature of our dwarf candidates, we show our derived color magnitude diagrams (CMDs) for each dwarf in Figure~\ref{fig:CMD}.  As expected from the spatial distributions seen in Figure~\ref{fig:diffuse}, only a handful of resolved point sources populate each CMD, consistent with a normal foreground stellar population and/or background compact galaxies. We can also easily simulate what our CMDs would have looked like had each dwarf actually been associated with M101, given their apparent magnitude in ground-based imaging and the CMDs of true M101 dwarfs observed with the same observational setup \citep{Bennet19}.  
These simulated CMDs were created using the {\it HST}-derived CMD from M101 DwA (M$_V$=$-$9.5; \citealt{Bennet19}), removing point sources at random until the total luminosity of the remaining sources is equal to that of the unresolved dwarf candidate, determined using the CFHTLS data and the distance of M101 DwA (D$=$6.83$^{+0.27}_{-0.26}$ Mpc).  In Figure~\ref{fig:CMD} we show the simulated CMD of what our brightest and faintest dwarf candidates would have looked like had they been associated with M101 -- in either case, a clear and well-populated red giant branch is apparent.  Given that these features are not observed, 
we conclude that Dw21, Dw22, Dw23 and Dw35 are background galaxies not associated with M101.

We measured the observational properties of our four diffuse dwarf galaxies in the {\it HST} data using GALFIT \citep{peng02} with a procedure identical to that presented in our previous work \citep{bennet17,Bennet19}, including inserting simulated diffuse dwarf galaxies to estimate our uncertainties \citep[see also][]{merritt14}.  We present these results in Table~\ref{tab:obj_unres}, alongside our ground-based CFHTLS measurements; {\it HST} F606W and F814W magnitudes were converted to the $V$-band using the relations of \citet{Sahu14}. There is good agreement between both datasets, although the {\it HST}-derived magnitude uncertainties tend to be slightly larger. 
Previous studies \citep{Merritt16,bennet17, Crnojevic19} have shown that the smaller primary mirror on {\it HST} is less effective at detecting the diffuse low surface brightness outer regions of unresolved targets. In addition, the small pixel scale of {\it HST} means that it is not optimised for unresolved LSB candidates. These factors combine so that in general the detected half-light radii for unresolved targets is smaller with {\it HST} than large ground based telescopes, with larger uncertainties both on the half-light radius and magnitude.

\section{The M101 Satellite Luminosity Function}\label{sec:LF}

As the four faint dwarf candidates that we have imaged were the only remaining viable members in the M101 sample of \citet{bennet17}, we can use their status as background objects to extend the M101 satellite luminosity function to fainter magnitudes.  As we discuss below, M101 is now only the third MW-sized halo with a near-complete luminosity function that pushes into the ultra-faint dwarf galaxy regime; after the MW \citep{mcconnachie12} and M31 \citep{martin16,McConnachie18}.  Such information is vital, as the number of such satellites is smaller than expected from dark matter only simulations \citep{moore99,klypin99}. Astrophysical mechanisms such as feedback, star formation efficiency and reionization may play a role in reconciling the differences between observed luminosity functions and cosmological simulations \citep{Brooks13,Sawala16,wetzel16,Garrison17, Simpson18}.  To ultimately solve this issue, however, the satellite luminosity function of many MW-sized halos should be measured so that we do not tune our results to the Local Group.

The dwarf galaxy candidates studied here (and in our previous work on the M101 luminosity function; \citealt{Bennet19}) were originally drawn from a semi-automated and well-characterized diffuse dwarf galaxy search of the CFHTLS \citep{bennet17}, utilizing information from the SBF-derived distances of subsequent work \citep{carlsten19}. It is important to know the limits of these studies when constructing our luminosity function.  First, the 50\% and 90\% completeness limits for identifying diffuse dwarfs in the CFHTLS with similar sizes to those in the Local Group are at $M_V$=$-$7.4 and $-$7.7 mag, respectively. 
Follow-up analysis using {\it HST} led to a nearly complete luminosity function to $M_V$=$-$8.2 mag, with the caveat that $\lesssim$3 true M101 dwarfs may have been missed as that program did not acquire complete {\it HST} imaging of dwarf candidates to that luminosity limit \citep{Bennet19}. Note that the SBF distance limits for the remaining dwarfs in that luminosity range suggest that none are M101 satellites \citep{carlsten19}.  

In the current work, we have presented {\it HST} imaging of the four dwarf candidates fainter than $M_V$=$-$8.2 mag which have SBF distance limits from \cite{carlsten19} consistent with M101 at the 1-$\sigma$ level.  Five other diffuse dwarf candidates with $M_V$$\gtrsim$$-$8.2 mag from the original CFHTLS sample remain -- Dw24, Dw25, Dw29, Dw36 and Dw37 -- but their formal distance limits do not overlap with M101 (although Dw24 does overlap at the $\sim$2-$\sigma$ level).  We note that none of these five dwarfs were targeted in the HI study of \citet{K20}, but future observations would be beneficial.  For the purposes of this work we assume that the SBF distance limits are correct and that these five dwarfs are not viable M101 members.  We are thus complete to the limit of the original CFHTLS diffuse dwarf search (with all of the caveats discussed above), corresponding to $M_V$=$-$7.7 ($-$7.4) mag at 90\% (50\%) completeness.

For our updated satellite luminosity function of M101, we include all galaxies reported within the projected virial radius of M101 ($\sim$250 kpc) and with a confirmed M101 distance using the TRGB method -- these dwarfs are listed in Table~3 of \citet{Bennet19}, although we have amended their absolute magnitudes to match our adopted distance of $D$=6.5 Mpc \citep{Beaton19}.  We do not include the bright dwarf UGC~08882 since its SBF distance places it slightly in the background of M101 \citep{Rekola05,carlsten19}.

\begin{figure*}
 \begin{center}
 \includegraphics[width=12cm]{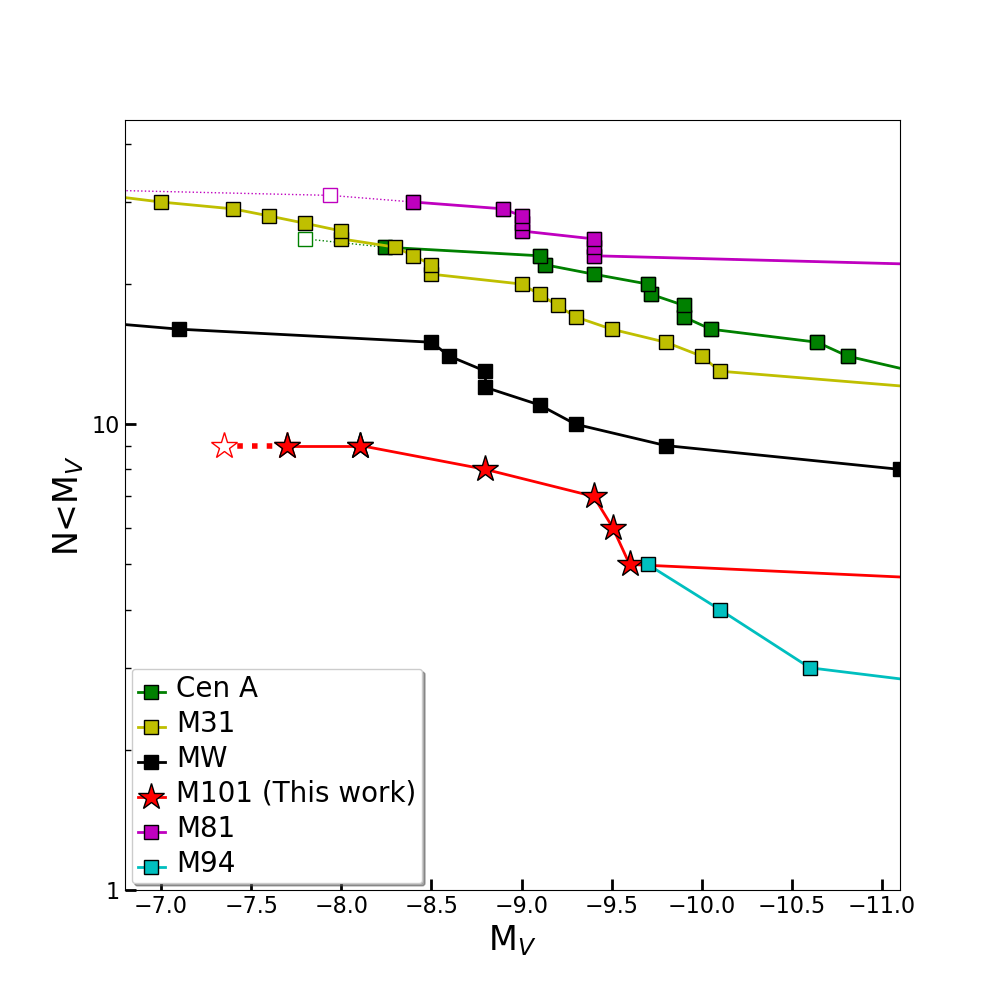}
 \caption{The cumulative satellite luminosity function for several Local Volume systems out to a projected radius of 250 kpc. 
We consider the M101 luminosity function to be 90\% complete down to $M_V$$\approx$$-$7.7 mag, and 50\% complete down to $M_V$$\approx$$-$7.4 mag (hollow symbol; we also mark the magnitude range between these two values with a dashed line); see Section~\ref{sec:LF} for details. The data for the other luminosity functons come from \cite{Smercina18} for M94, \cite{Crnojevic19} for Cen A, 
 \cite{chiboucas13} and \cite{Smercina17} for M81, \cite{martin16} \&  \cite{McConnachie18} for M31 and \cite{mcconnachie12} for the MW. Note that this is a lower limit for the MW due to incomplete spatial coverage; no attempt was made to correct any luminosity function for incompleteness. We denote the region where the Cen A and M81 luminosity functions become incomplete with hollow symbols and dashed lines, as reported by \citet{Crnojevic19} and \citet{chiboucas13}, respectively. Galaxies are listed in descending order of stellar mass.  \label{fig:LF_NGC}}
 \end{center}
\end{figure*}

The cumulative satellite luminosity function for M101 is shown in Figure~\ref{fig:LF_NGC}, along with those of several other Local Volume systems: the MW \citep[][and references therein]{mcconnachie12}, M31 \citep{martin16,McConnachie18}, M81 \citep{chiboucas13,Smercina17}, M94 \citep{Smercina18} and Cen~A \citep{Crnojevic19}.  These galaxies span a narrow range of total masses ($\sim$2.5--9$\times$10$^{11}$ M$_{\odot}$, based on globular cluster dynamics within 40 kpc where available; \citealt{Eadie16}, \citealt{Woodley10}) and illustrate the range in satellite properties amongst the sample.  In the figure, we mark the extension fainter than $M_V$=$-$7.7 to denote where the M101 luminosity function becomes significantly incomplete. We also mark the approximate completeness limits for the luminosity function of Cen~A \citep{Crnojevic19} and M81 \citep{chiboucas13}, which are at $M_V$$\approx$$-$ 8.0 and $M_V$$\approx$$-$8.1, respectively.  None of the reported luminosity functions have been corrected for incompleteness effects; in particular, the MW luminosity function in practice is a lower limit, as our ability to detect satellites near the Galactic plane is limited, but it is well quantified \citep[e.g.][]{Drlica-Wagner19}.  To our knowledge, M101 is only the third MW-like galaxy with a well-measured satellite luminosity function that extends into the ultra-faint dwarf galaxy regime.

At the bright end ($M_V$$\lesssim$$-$14 mag, which is not displayed in Figure~\ref{fig:LF_NGC}), the M101 luminosity function is similar to that of the MW and M31, along with most of the other Local Volume sample.  However, at faint magnitudes, M101 has significantly fewer satellites, and the fact that no new M101 satellites were identified in the current work between $-$7.4 and $-$8.2 mag exacerbates the differences with the Local Group.  For instance, M101 has five satellites with $-$14$<$$M_V$$<$$-$7.7 mag, while the MW has eleven.  Overall, M101 has only eight satellites brighter than $M_V$=$-$7.7 while the MW and M31 have 14 and 26, respectively -- a factor of $\sim$3 scatter between systems.  One other Local Volume galaxy, M94, also has a deficit of satellites, but is only complete to $M_V$$\approx$$-$9.1 \citep{Smercina18}; future observations should probe fainter satellites around this and other systems to further understand the scatter in MW-like hosts.

\section{Discussion and Conclusions} \label{sec:conclusion}

We have presented {\it HST} follow-up imaging of four potential M101 dwarf satellite galaxies (Dw21, Dw22, Dw23, and Dw35 from \citealt{bennet17}), extending the M101 luminosity function into the ultra-faint dwarf galaxy regime.  In all four cases, the {\it HST} imaging displays unresolved diffuse emission, consistent with a galaxy at a much larger distance than M101.  To further establish the background nature of these objects, we generated simulated CMDs of dwarfs at the M101 distance; they  
clearly demonstrate that each dwarf would have displayed a clear, resolved red giant branch if it were associated with M101 
(Figure~\ref{fig:CMD}).  

One hallmark of this work is that the dwarf candidates come from a well-quantified search for M101 dwarfs with the CFHTLS  
\citep{bennet17}.  This, combined with SBF-derived distance estimates of the same data set \citep{carlsten19}, allowed us to calculate an extended satellite luminosity function for M101 which is 90\% complete at $M_V$=$-$7.7 mag and 50\% complete at $M_V$=$-$7.4 mag, thus dipping into the ultra-faint dwarf galaxy regime. We confirm that M101 is very deficient in faint satellites, with only eight systems with $M_V$$\lesssim$$-$7.7 mag, compared to the 14 and 26 around the Milky and M31, respectively.  A systematic and rigorous observational census of dwarf galaxies around MW-like systems is warranted to understand the overall demographics of satellites, with galaxy environment being a potential driver of any trends \citep[][]{Bennet19}.  It has also been suggested that the bulge-to-total baryonic mass ratio is an indicator of satellite number in MW analogs \citep{Java20}.

The targets chosen for the present study were the four remaining dwarf candidates that have SBF-derived distances consistent with M101 \citep{carlsten19}. While we have shown that none of these four are actually related to M101, their SBF distance estimates are highly uncertain due to their faintness, which pushes the SBF technique to its limit, which seems to correspond to $g$$\sim$21 mag in the CFHTLS data set (the magnitude of Dw9, the faintest true M101 member identified by the SBF technique).  Additionally, atomic hydrogen (HI) observations are another vital technique for screening dwarfs, and for probing the astrophysics of gas stripping and quenching, as recently demonstrated around M101 \citep[][although the dwarf candidates in the present study were below the brightness cutoff of that work]{K20}. Indeed, for clumpy and/or star-forming dwarf galaxy targets, HI may be the only reliable means of screening or estimating distances prior to {\it HST} imaging, as the SBF technique is not appropriate in these circumstances.  For instance, for the clumpy M101 dwarf candidate dw1408+56, the SBF technique estimated a distance of $D$$\approx$12 Mpc \citep{carlsten19}, while HI observations showed that this dwarf was actually at a significantly larger distance \citep[$V_{sys}$=1904 km s$^{-1}$ or $D_{HI}$=27 Mpc;][]{K20}. 
We conclude that both ground-based SBF distance estimates and HI observations should be used to guide deeper follow-up studies of dwarf systems in the Local Volume when appropriate, but that {\it HST}-quality data is a necessity for ultimately measuring satellite luminosity functions.  These observations are vital for further testing the $\Lambda$CDM model on small scales.

It will be possible to go even further down the satellite luminosity function of M101 and other systems in the Local Volume with future wide-field space missions such as the {\it Wide Field Infrared Survey Telescope} \citep[{\it WFIRST};][]{wfirst}.  As can be seen from Figure~\ref{fig:CMD}, dwarfs as faint as $M_V$$\approx$$-$7 mag should be detectable with moderate exposure times of $\sim$1 hour.  Further, the {\it WFIRST} Wide Field Instrument will have a field of view of 0.281 deg$^2$, making a dwarf search out to $\sim$250 kpc (similar to that done by \citealt{bennet17} and the CFHTLS) possible in $\approx$32 pointings, or 32 hours of exposure time.
An ambitious {\it WFIRST} program such as this would make it possible to measure the satellite luminosity function into the ultra-faint dwarf galaxy regime throughout the Local Volume in the decade to come.

\acknowledgments

We are grateful to the referee for a careful reading of the manuscript and for his/her useful suggestions that helped improve this work.

Based on observations with the NASA/ESA Hubble Space Telescope, which is operated by the Association of Universities for
Research in Astronomy, Incorporated, under NASA contract NAS5-
26555. Support for Program number HST-GO-15858.003-A was
provided through a grant from the STScI under NASA contract NAS5-
26555.

Research by PB is supported by NASA through grant number HST-GO-14796.005-A from the Space Telescope Science Institute which is operated by AURA, Inc., under NASA contract NAS 5-26555. Research by DJS is supported by NSF grants AST-1821967, 1821987, 1813708, 1813466, and 1908972.   Research by DC is supported by NSF grant AST-1814208, and by NASA through grants number HST-GO-15426.007-A and HST-GO-15332.004-A from the Space Telescope Science Institute, which is operated by AURA, Inc., under NASA contract NAS 5-26555. KS acknowledges support from the Natural Sciences and Engineering Council of Canada (NSERC).

\vspace{5mm}
\facilities{Hubble Space Telescope}

\software{Astropy \citep{astropy13,astropy18}, DOLPHOT v2.0 photometric package \citep{dolphin00}, SExtractor \citep{bertin96}, GALFIT \citep{peng02}
}

\bibliographystyle{aasjournal}
\bibliography{ref_PB}

\end{document}